 \documentstyle{article}       
\begin{document}
\begin{titlepage}
\title{On maximum number of decoherent histories}
\author{Lajos Di\'osi
\thanks{E-mail: diosi@rmki.kfki.hu}\\
KFKI Research Institute for Particle and Nuclear Physics\\
H-1525 Budapest 114, POB 49, Hungary\\\\
{\it bulletin board ref.: gr-qc/9409028}}
\date{}
\maketitle
\begin{abstract}
It is shown that $N^2$ is the upper limit for the number of histories in a
decohering family of $N$-state quantum system. Simple criterion is
found for a family of $N^2$ fine grained decohering histories
of Gell-Mann and Hartle to be identical with a family of Griffiths'
consistent quantum trajectories.
\end{abstract}
\end{titlepage}

Several versions of consistent history theories
\cite{Gri84,GMH93,Omn92,DowHal92,Gri93} have been
proposed earlier to incorporate the interpretation of quantum state into
the basic equations of the formalism. Here we are going to prove that
the maximum number of consistent histories is equal in both
decoherent history and consistent quantum trajectory theories of Gell-Mann
and Hartle \cite{GMH93}, and of Griffiths \cite{Gri93}, respectively.
Further numerous algebraic properties have been explored independently by
Dowker and Kent \cite{DowKen94}.

Let us consider a closed N-state quantum system whose state space is the
Hilbert space ${\cal H}$ of N dimensions. Let us introduce a time-ordered
sequence of events and the corresponding operator
\begin{equation}
C_\alpha\equiv P_{\alpha_n}(t_n)\dots P_{\alpha_2}(t_2)P_{\alpha_1}(t_1)
\end{equation}
where the $P_{\alpha_k}(t_k)$'s form complete orthogonal sets of Hermitian
projectors for each $k=0,1,\dots,n$ in turn. Given the Heisenberg state
$\rho$ of the system, the $C_\alpha$'s are said to generate a family of
{\it decoherent histories} \cite{GMH93}
provided the so-called {\it decoherence fuctional} is diagonal:
\begin{equation}
D(\alpha,\beta)\equiv tr\left(C_\beta^\dagger C_\alpha\rho\right)
        =0~~~~for~all~\alpha\neq\beta.
\end{equation}
Its diagonal elements will be assigned to decoherent
histories as their probabilities $p$:
\begin{equation}
p(\alpha)=D(\alpha,\alpha).
\end{equation}

Griffiths' quantum trajectories \cite{Gri93}
can be identified by a certain subset of {\it fine grained} \cite{GMH93}
decoherent histories. Accordingly, all events in the sequence (1) will
be described by 1-dimensional (pure state) projectors.
Griffiths assignes probabilities to his quantum trajectories
without referring to the state $\rho$ and the decoherence functional either.
Here, nevertheless, we show how Griffiths noninterference condition
and his probability assignment can be formulated within the context of
decoherent histories (1-3).
To this end, assume that the decoherence condition (2) holds  for a family of
fine grained histories, i.e., the decoherence functional is diagonal.
Griffiths noninterference condition imposes additional constraints on the
nonzero diagonal elements. For any pair of initial and
final labels $\alpha_1,\alpha_n$, {\it at most one} decoherent history
$\alpha_1,\alpha_2,\dots,\alpha_k\dots,\alpha_n$ should make the
corresponding diagonal
element $D(\alpha,\alpha)$ nonvanishing. In other words, the endpoints
$\alpha_1,\alpha_n$ should determine the whole string uniquely.
The following factorization is then valid for nonzero diagonal elements:
\begin{eqnarray}
D(\alpha,\alpha)
=tr\Bigl(P_{\alpha_n}(t_n)P_{\alpha_{n-1}}(t_{n-1})\Bigr)
 tr\Bigl(P_{\alpha_{n-1}}(t_{n-1})P_{\alpha_{n-2}}(t_{n-2})\Bigr)\dots
 \nonumber\\
 \dots\times tr\Bigl(P_{\alpha_2}(t_2)P_{\alpha_1}(t_1)\Bigr)
 tr\Bigl(P_{\alpha_1}(t_n)\rho\Bigr).
\end{eqnarray}
Such family is considered consistent \`a la Griffiths,
and assigned by the following probabilities:
\begin{equation}
p^G(\alpha)={p(\alpha)\over p(\alpha_1)}.
\end{equation}
where $p(\alpha_1)=tr\left(P_{\alpha_1}(t_1)\rho\right)$.
{}From Eq.~(4) it follows that Griffiths' probabilities $p^G$ (5) are
independent of the Heisenberg state $\rho$ and depend only
on the sequence (1) of events.
Griffiths probability (5) of a given trajectory $\alpha$
is equal to the corresponding GMH probability (3) at the condition that
the earliest event was $P_{\alpha_1}(t_1)$ with certainty; this condition
relaxes the $\rho$-dependence of the Griffiths probabilities.

Since the total number of pairs $\alpha_1,\alpha_n$ is $N^2$,
the maximum number of Griffiths' trajectories in a consistent
family is $N^2$. In the generic case, the number of Griffiths' trajectories
is equal to the number of pairs $\alpha_1,\alpha_n$ for which
the overlap of the corresponding initial and final states is nonzero, i.e.:
\begin{equation}
tr\Bigl(P_{\alpha_n}(t_n)P_{\alpha_1}(t_1)\Bigr)
\neq 0.
\end{equation}

We show that $N^2$ is the upper limit for the number of decohering
histories in GMH families, too. Let us start initially with pure state
$\rho=|\psi\rangle\langle\psi|$ and introduce the {\it unnormalized} vectors
\begin{equation}
|\varphi_\alpha\rangle=C_\alpha|\psi\rangle.
\end{equation}
{}From the decoherence condition (2) it follows that they are orthogonal
to each other:
\begin{equation}
\langle\varphi_\beta|\varphi_\alpha\rangle=0~~~~for~all~\alpha\neq\beta.
\end{equation}
The maximum number of (nonzero) orthogonal vectors in ${\cal H}$ is $N$.

Let us allow mixed states, too. Consider the orthogonal expansion
$\rho=\sum_{r=1}^N w_r |\psi_r\rangle\langle\psi_r|$ which is always
possible with nonnegative normalized wights $w_r$.
Consider the trivial embedding of our system into a larger one whose
state space is  ${\cal H}\otimes{\cal H}^\prime$ where ${\cal H}^\prime$
is also $N$-dimensional Hilbert space.
Construct the following state vector in ${\cal H}\otimes{\cal H}^\prime$:
\begin{equation}
|\Psi\rangle=
\sum_{r=1}^N \sqrt{w_r}|\psi_r\rangle\otimes|\psi_r^\prime\rangle
\end{equation}
where $\{|\psi_r^\prime\rangle,r=1,\dots,N\}$ form a complete orthonormal
system in ${\cal H}^\prime$. Introduce the following {\it unnormalized}
vectors in ${\cal H}\otimes{\cal H}^\prime$:
\begin{equation}
|\Phi_\alpha\rangle=\left(C_\alpha\otimes{\bf 1}\right)|\Psi\rangle.
\end{equation}
{}From Eqs.~(7-10) one can prove that these vectors are also orthogonal to each
other:
\begin{equation}
\langle\Phi_\beta|\Phi_\alpha\rangle=\langle\varphi_\beta|\varphi_\alpha\rangle
=0~~~~for~all~\alpha\neq\beta.
\end{equation}
In Hilbert space ${\cal H}\otimes{\cal H}^\prime$, the maximum number of
orthogonal vectors is equal to the number $N\times N$ of dimensions.
In such a way we have proven that the
maximum number of histories in a given decohering family is $N^2$.
(The general limit is $rank(\rho)N$, as one could easily prove.)

We see that the maximum number of consistent histories is identical
at both GMH and Griffiths conditions. This fact makes one think about
the extent of GMH decohering families are more general than Griffiths' ones.
We are not yet able to present an exhaustive mathematical comparison.
Nevertheless, we know the
class of GMH families which is identical to Griffiths' families.
If {\it all} initial events $P_{\alpha_1}(t_1)$ has
nonvanishing overlaps (6) with {\it all} final events
$P_{\alpha_n}(t_n)$ then a family of fine grained decoherent histories
is a family of Griffiths' quantum trajectories.

The proof is rather easy. From the existence of all overlaps (6) it follows
that, for each pair $\alpha_1,\alpha_2$ there must be at least one
decoherent history $\alpha_1,\dots,\alpha_k,\dots,\alpha_n$
such that $D(\alpha,\alpha)$ is nonzero. [Otherwise $D(\alpha,\alpha)$ would
be zero for all $\alpha_2\dots\alpha_{n-1}$ and this would contradict to (6).]
Such decoherent histories must exist for
all $N^2$  different pairs $\alpha_1,\alpha_n$. Since the maximum number
of decoherent histories, as shown above, is also $N^2$
there can be only a single one for each fixed
pair $\alpha_1,\alpha_n$. This means that Griffiths' additional condition
fulfils for the given family of $N^2$ GMH decoherent histories.

For the generic fine grained decohering family when the number of
positive overlaps (6) may be less than $N^2$ it is
most likely that multiple histories will arise
with the same endpoints $\alpha_1,\alpha_n$ and this denies the equivalence
with any Griffiths family. We can not judge whether Griffiths' conditions
are too restrictive or, oppositely, the GMH decoherence is too general.
Nevertheless, GMH decoherence theory is {\it formally} equivalent
with the von Neumann collapse theory \cite{Dio92}.
Any restriction of the GMH decoherence might thus miss
the probability assignment for certain families which
are otherwise well-interpreted in the von Neumann language.

\bigskip
This work was supported by the grant OTKA No. 1822/1991.



\begin{thebibliography}{99}
\bibitem{Gri84} R.B. Griffiths, J. Stat. Phys. 36 (1984) 219.
\bibitem{Omn92} R. Omn\`es, Rev. Mod. Phys. 64 (1992) 339.
\bibitem{GMH93} M. Gell-Mann and J.B. Hartle, Phys. Rev. D 47 (1993) 3345.
\bibitem{DowHal92}H.F. Dowker and J.J. Halliwell, Phys. Rev. D 46 (1992) 1580.
\bibitem{Gri93} R.B. Griffiths, Phys. Rev. Lett. 70 (1993) 2201.
\bibitem{DowKen94} F. Dowker and A. Kent, preprint gr-qc/9409037;
        Cambridge preprint DAMTP/94-48.
\bibitem{Dio92} L. Di\'osi, Phys. Lett. 280 B (1992) 71.
\end{thebibliography}
\end{document}